\documentclass[journal=jctcce,manuscript=article,keywords=true]{achemso}

\usepackage[version=3]{mhchem} 
\usepackage{hyperref}
\usepackage{xcolor}
\usepackage{soul}
\usepackage[normalem]{ulem}
\sethlcolor{yellow}

\setkeys{acs}{keywords = true}

\author{Susanta Das}
\affiliation{Computational Life Sciences, The Cleveland Clinic Research, The Cleveland Clinic, Cleveland, Ohio 44195, USA}

\author{Subhamoy Bhowmik}
\affiliation{Computational Life Sciences, The Cleveland Clinic Research, The Cleveland Clinic, Cleveland, Ohio 44195, USA}
\alsoaffiliation{Department of Chemistry, Michigan State University, East Lansing, Michigan 48824, USA}

\author{Zhen Li}
\affiliation{Computational Life Sciences, The Cleveland Clinic Research, The Cleveland Clinic, Cleveland, Ohio 44195, USA}

\author{Milana Bazayeva}
\affiliation{Computational Life Sciences, The Cleveland Clinic Research, The Cleveland Clinic, Cleveland, Ohio 44195, USA}

\author{Danil Kaliakin}
\affiliation{Computational Life Sciences, The Cleveland Clinic Research, The Cleveland Clinic, Cleveland, Ohio 44195, USA}

\author{Akhil Shajan}
\affiliation{Computational Life Sciences, The Cleveland Clinic Research, The Cleveland Clinic, Cleveland, Ohio 44195, USA}

\author{Kenneth M. Merz Jr.}
\email{merzk@ccf.org}
\affiliation{Computational Life Sciences, The Cleveland Clinic Research, The Cleveland Clinic, Cleveland, Ohio 44195, USA}
\alsoaffiliation{Department of Chemistry, Michigan State University, East Lansing, Michigan 48824, USA}

\title
{Quantum Computing Enabled \emph{ab initio} Molecular Dynamics Simulations}


\begin{document}

\begin{tocentry}

\centering
\includegraphics[width=\linewidth]{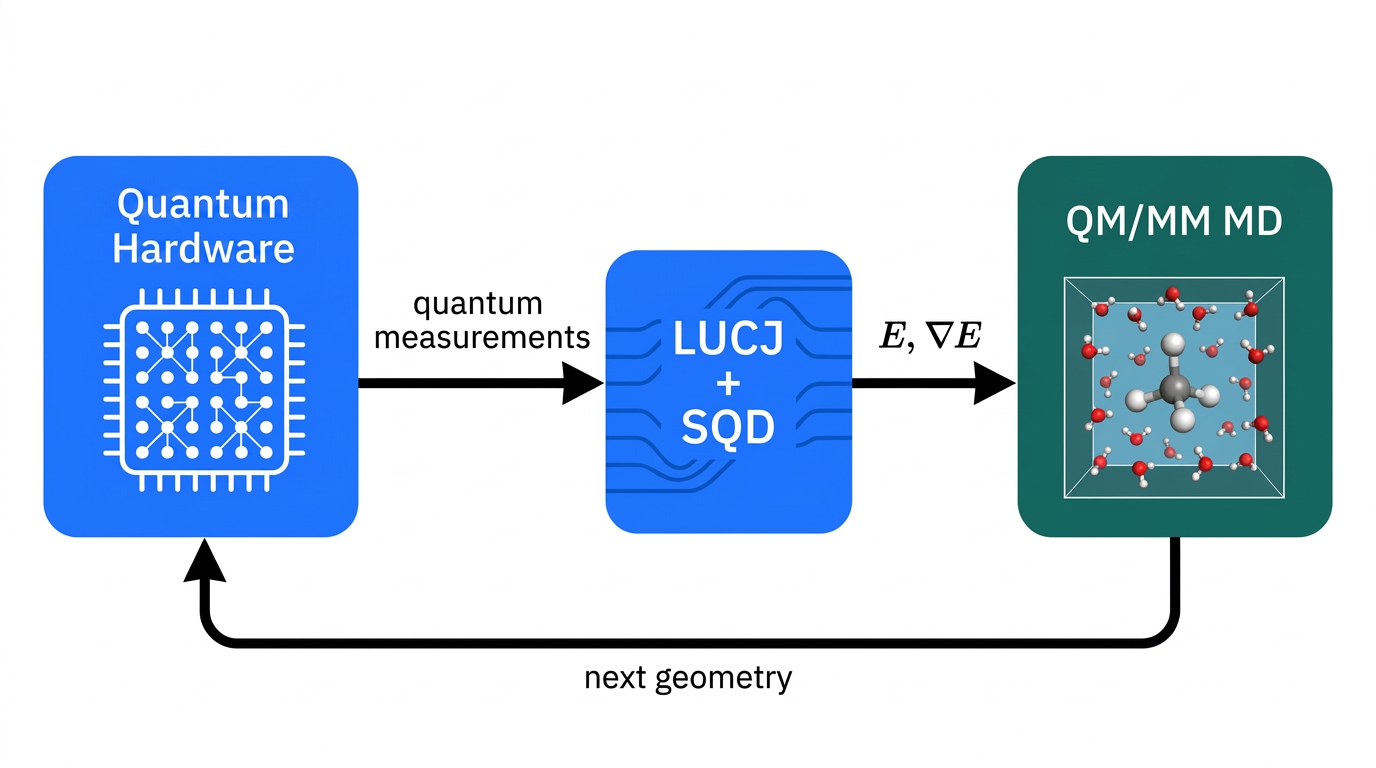}
\end{tocentry}

\begin{abstract}
We demonstrate a quantum-classical workflow for \emph{ab initio} molecular dynamics (AIMD) in which quantum measurements from a chemistry-inspired LUCJ ansatz are post-processed using Sample-based Quantum Diagonalization (SQD) to recover determinant subspaces and deliver energies and \emph{analytical} nuclear gradients for dynamics. As an exact benchmark, we use full configuration interaction (FCI) in the STO-3G basis, enabling a direct assessment of the accuracy of SQD. In gas-phase benchmarks, SQD reproduces FCI energies and gradients to within 1 kcal mol$^{-1}$ of the FCI reference and yields stable AIMD trajectories. In explicit-solvent QM/MM simulations, SQD retains this agreement, matching FCI energy fluctuations and RMS gradient profiles and reproducing solute--solvent structure as quantified by radial distribution functions. Overall, these benchmarks establish LUCJ+SQD as a practical route for integrating current quantum hardware into QM/MM molecular dynamics and provide an early demonstration of condensed-phase QM/MM dynamics driven by a quantum electronic-structure engine.
\end{abstract}

\section{Introduction}

\emph{Ab initio} molecular dynamics (AIMD) has become a central tool for studying chemical and biological processes in the condensed phase, since it propagates nuclei on potential energy surfaces obtained directly from quantum mechanical electronic structure calculations.\cite{car1985unified,marx2009ab,hassanali2014aqueous,sakti2020recent} AIMD underpins applications ranging from liquid structure and vibrational spectroscopy to proton transfer, ion solvation, and drug binding in complex environments.\cite{sakti2020recent, iftimie2005ab, thomas2013computing, guo2023ab,xi2022ion, qian2023ab, mirza2025calcite, wang2023aimd} In practice, however, AIMD is often limited to density functional theory (DFT) or low-order wave function methods, and to relatively modest system sizes and time scales, because the electronic structure problem must be solved repeatedly along a trajectory. These limitations are particularly severe when accurate treatment of noncovalent interactions, charge transfer, and polarization in water and biomolecular environments is required.\cite{hobza2016introduction, wang2024ab, schade2022towards, worner2017charge, bedrov2019molecular}

Classical molecular dynamics with empirical force fields alleviates much of the cost and remains the workhorse for long time scale simulations of proteins, nucleic acids, and materials.\cite{ponder2003force, mackerell2004empirical, maier2015ff14sb,vanommeslaeghe2010charmm,doherty2017revisiting, christen2005gromos} Force fields have been refined for decades using high-level quantum chemistry and experimental data, and modern models can describe many classes of systems with good accuracy.\cite{ding2023data, frohlking2020toward, chipot2024recent, reith2011modern, ringrose2022exploration} Still, they struggle in situations that involve significant electronic rearrangement, unusual chemotypes, or strongly non-additive interactions, and they inherit any deficiencies of the underlying quantum data used during parametrization.\cite{harrison2018review, heindel2023many, he2022recent, unke2021machine, seiferth2023limitations} These issues are especially acute for solvation and binding processes, where subtle balance between hydrogen bonding, dispersion, electrostatics in water and heterogeneous environments control the thermodynamics.\cite{mobley2009small, mobley2012perspective, mobley2014freesolv, slochower2019binding}

Hybrid quantum mechanics/molecular mechanics (QM/MM) approaches provide a compromise by treating a chemically active region at a quantum mechanical level while embedding it in a classical environment.\cite{jorgensen2013foundations, senn2009qm, lin2024qm, li2024molecular} QM/MM AIMD has been widely used to study enzyme mechanisms, photochemistry, and ion solvation, and has been incorporated into workflows for free energy and binding affinity calculations, including book-ending and alchemical free energy (AFE) schemes.\cite{lonsdale2012practical, naray2013quantum, yang2010qm, magalhaes2020modelling, senn2007qm, monard1999combined, song2020evolution} Recent work from our group has shown that book-ending corrections computed at a QM/MM level can systematically improve hydration free energies and binding thermodynamics when combined with MM-based sampling,\cite{bazayeva2025quantum} and that coupling \textsc{Amber} to \textsc{Quick} via \texttt{sander} allows Hartree-Fock (HF), DFT, and full configuration interaction (FCI) calculations for the QM region within a single workflow.\cite{cruzeiro2021open, manathunga2023quantum, case2023ambertools, rahnamoun2020reaxff, bazayeva2025quantum} Extending this interface to external CI solvers, including \textsc{PySCF}-based FCI and Sample-based Quantum Diagonalization (SQD), has already enabled quantum-centric alchemical free energy calculations for small solutes in water and demonstrated that quantum hardware can be embedded into classical AFE pipelines.\cite{bazayeva2025quantum}

Quantum computing offers a conceptually different way to tackle electronic structure problems by exploiting superposition and entanglement to represent many-electron wave functions more compactly than on classical hardware.\cite{aspuru2005simulated, kassal2011simulating, cao2019quantum, weidman2024quantum} Within the quantum-centric computing paradigm, classical high-performance computing (HPC) resources orchestrate quantum subroutines and perform heavy classical post-processing, while quantum processing units (QPUs) are used selectively for tasks that are classically expensive.\cite{robledo2025chemistry, alexeev2024quantum} QSCI and SQD are key examples of this approach, in which a parameterized quantum circuit is used to sample important determinants, and classical diagonalization in the sampled subspace recovers approximate CI energies.\cite{robledo2025chemistry, kanno2023quantum, nakagawa2024adapt} SQD augments QSCI with configuration recovery and carryover procedures that improve sampling robustness on noisy hardware and has already been validated across a range of chemical problems, including metal complexes, aromatic systems, open-shell and excited-state cases, implicit solvent, QM/MM test systems, and thermochemistry datasets.\cite{robledo2025chemistry, kaliakin2025implicit, kaliakin2025accurate, shajan2025toward} Quantum-centric simulations of the water and methane dimers have shown that SQD can reproduce FCI and selected-CI potential energy surfaces for hydrogen-bonded and dispersion-bound complexes with that agree with classical solvers, using circuits in the 27--54 qubit range on superconducting hardware.\cite{robledo2025chemistry, kaliakin2025implicit} A recent perspective maps scientifically meaningful quantum-chemistry use cases to the first fault-tolerant window of roughly 25--100 logical qubits, highlighting opportunities such as phase estimation, real-time dynamics, and active-space embedding.\cite{alexeev2025perspective, verma2025multireference}

In parallel with these developments, several works have begun to explore quantum algorithms for molecular dynamics and finite-temperature properties. Fedorov and co-workers introduced a VQE-based AIMD scheme in which the electronic ground state at each geometry is obtained variationally on a quantum computer and forces are estimated through correlated sampling and finite differences.\cite{fedorov2021ab} That work demonstrated a proof-of-principle quantum AIMD trajectory for \ce{H2} on NISQ hardware and highlighted both the promise and the challenges associated with force estimation, noise, and circuit depth. Other groups have proposed related VQE-based dynamics methods, hybrid schemes for Hessian and vibrational analysis, and quantum algorithms for free energy estimation,\cite{hirai2023excited, ollitrault2021molecular, mishima2008quantum, bassman2022computing, paudel2022quantum} but these studies have largely been confined to simple molecules in the gas-phase, very short trajectories, or classical circuit simulators. 

In the present work, we extend the SQD-based quantum-centric framework from static calculations to fully dynamical simulations and present an early demonstration of condensed-phase QM/MM \emph{ab initio} molecular dynamics driven directly by a quantum electronic-structure solver. We develop an interface in which \textsc{Amber}'s \texttt{sander} module performs QM/MM molecular dynamics, \textsc{Quick} provides integral evaluation for a QM region, \textsc{PySCF} constructs a FCI Hamiltonian in a minimal basis, and SQD running on superconducting hardware supplies approximate FCI energies that are used to compute forces along the trajectory.\cite{bazayeva2025quantum} This setup enables a direct comparison between FCI-based and SQD-based AIMD for the same QM/MM Hamiltonian. We focus on three prototypical solutes, \ce{NH3}, \ce{CH4}, and \ce{H2O}, in explicit liquid water, as well as on gas-phase benchmark trajectories, and assess the performance of SQD in terms of energies, forces, and solute--solvent radial distribution functions.

\begin{figure}
  \centering
  \includegraphics[width=\linewidth]{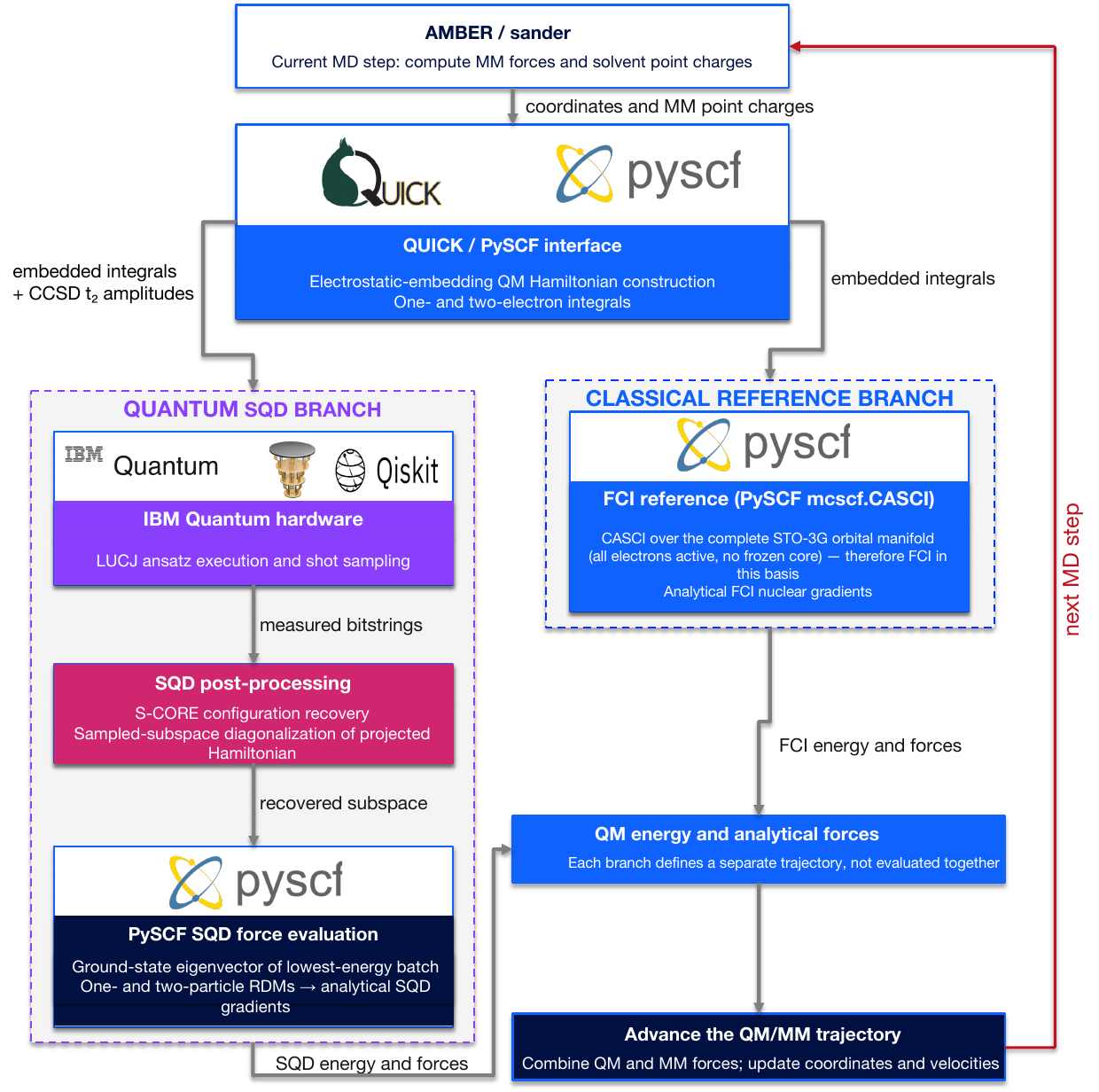}
  \caption{Schematic of the SQD-enabled QM/MM \emph{ab initio} molecular dynamics workflow integrating \textsc{Amber}, \textsc{Quick}, \textsc{PySCF}, and IBM quantum hardware. At each MD step, \textsc{Amber}/\texttt{sander} supplies the solute coordinates and MM point charges, from which the \textsc{Quick}/\textsc{PySCF} interface builds the electrostatically embedded QM Hamiltonian. The classical reference branch uses the \textsc{PySCF} \texttt{mcscf.CASCI} kernel over the complete STO-3G orbital manifold, with all electrons active and no frozen core, and is therefore FCI in this basis. The quantum branch executes a LUCJ ansatz on \texttt{ibm\_cleveland} and post-processes the measured bitstrings by SQD. Both branches yield \emph{analytical} nuclear gradients and define separate trajectories, run independently for comparison. The QM energy and forces are returned to \textsc{Amber} and combined with the MM forces to advance the trajectory. Protocol details are given in Sections 2.5--2.8 and the Supporting Information.}
  \label{fig:workflow_sqd_aimd}
\end{figure}

SQD-driven AIMD can be viewed as a dynamical counterpart to recent quantum-centric studies of free energies and noncovalent interactions, and as a concrete step toward practical quantum AIMD in chemically relevant environments.\cite{bazayeva2025quantum, kaliakin2025implicit, kaliakin2025accurate, shajan2025toward, shajan2025molecular} By coupling SQD to a standard QM/MM molecular dynamics engine, current quantum hardware can already be integrated into condensed-phase workflows and can reproduce FCI reference energies and forces to within 1 kcal mol$^{-1}$ of the FCI reference for small solutes in water. At the same time, these results highlight the main bottlenecks for scaling, including measurement noise in force estimation, active-space size, and hardware limitations, motivating future extensions to larger solutes and longer trajectories. A key feature of SQD in this context is its ability to compress the full active-space Hilbert space to a much smaller determinant subspace selected from hardware measurement outcomes, while retaining near-FCI accuracy for the systems studied here. For \ce{CH4}(aq) this corresponds to diagonalizing in roughly 5\% of the 15{,}876-determinant STO-3G configuration space (Supporting Information, Table~S2). The complete workflow is summarized in Figure~\ref{fig:workflow_sqd_aimd}.

\section{Computational Details}

\subsection{Systems}

All molecular dynamics simulations were carried out on neutral solutes in the gas phase and in explicit water. Ammonia (\ce{NH3}) was used as the primary test system, and methane (\ce{CH4}) and water (\ce{H2O}) were employed as additional production systems in solution. The gas-phase simulations involve a single \ce{NH3} molecule in vacuum. The condensed-phase simulations consist of a single solute molecule (\ce{NH3}, \ce{CH4}, or \ce{H2O}) in a periodic box of classical water molecules. Specific box parameters and water model details are provided in the Supporting Information.

\subsection{Electronic Structure and Active Spaces}
\label{subsec:activesp}

For all SQD and reference calculations, we employed the STO-3G basis. The QM region was treated with active spaces of (10e,8o) for \ce{NH3}, (10e,9o) for \ce{CH4}, and (10e,7o) for \ce{H2O}. The classical reference is FCI in the same active space, which is therefore full configuration interaction in the STO-3G basis and provides an exact benchmark for SQD subspace recovery.

The choice of the STO-3G minimal basis was deliberate. It produces sufficiently small active-space Hamiltonians to be mapped onto contemporary 16--21-qubit hardware while still presenting a non-trivial electron-correlation problem and giving access to an exact in-active-space reference for benchmarking SQD subspace recovery.

For each solute, \emph{all electrons} were treated as active and the active space comprised the \emph{full} STO-3G molecular-orbital manifold, with no frozen core and no orbital truncation: this yields all-electron active spaces of (10e,8o) for \ce{NH3}, (10e,9o) for \ce{CH4}, and (10e,7o) for \ce{H2O}, which together cover every electron and every spatial orbital available in the minimal basis. The reference is therefore full configuration interaction in the STO-3G basis; because the CI space is complete, it is invariant to rotation of the underlying Hartree--Fock orbitals. It was executed with PySCF's \texttt{mcscf.CASCI} driver, which reduces to FCI when the active space spans the complete orbital manifold, together with the corresponding analytical nuclear gradients. The corresponding $M_s = 0$ determinant spaces comprise 3{,}136, 15{,}876, and 441 configurations for \ce{NH3}, \ce{CH4}, and \ce{H2O}, respectively; a single SQD batch of $N = 200$--800 bitstring samples therefore addresses 1.3--5.0\% of the \ce{CH4} space, 6.4--25.5\% of the \ce{NH3} space, and 45--100\% of the much smaller \ce{H2O} space (Supporting Information, Table~S2).

All one- and two-electron integrals and analytical FCI gradients were generated with PySCF.\cite{sun2018pyscf}

Condensed-phase simulations were performed in explicit OPC water.\cite{izadi2014building, izadi2016accuracy, xiong2020melting} The QM/MM implementation is based on \textsc{Amber}/\texttt{sander} using electrostatic embedding, where QM energies and forces are supplied either by FCI or by SQD as described in Section~\ref{subsec:sqd}.

\subsection{QM/MM Embedding}
\label{subsec:qmmmem}

Condensed-phase simulations were performed using an electrostatically embedded QM/MM scheme in which the solute is treated quantum mechanically and the solvent is described by a classical force field. The systems were solvated using the OPC water model. In the electrostatic embedding, MM point charges enter the QM Hamiltonian as an external potential, and QM--MM nonbonded interactions are described by Coulomb and Lennard--Jones terms. The QM region description is identical in the classical reference calculations (FCI in the chosen active space and basis (see Section~\ref{subsec:activesp}) and in the SQD-driven simulations, so that any differences arise only from the electronic-structure engine used to obtain the QM energies and forces.

The QM/MM implementation is based on \textsc{Amber} (\texttt{sander}) coupled to \textsc{Quick}, using a \textsc{Quick}/PySCF interface for the embedded QM calculations. In SQD-driven simulations, the same QM/MM embedding framework is retained, with the QM energies provided by SQD as described in Section~\ref{subsec:sqd}.

\subsection{Sample-based Quantum Diagonalization}
\label{subsec:sqd}

Sample-based quantum diagonalization (SQD) is used to obtain active-space electronic energies from a quantum processor while retaining a classical diagonalization step in a compact many-electron subspace. For each molecular geometry, we define an active-space electronic Hamiltonian in second quantization,
\begin{equation}
\hat H \;=\; E_0 \;+\; \sum_{pq} h_{pq}\,\hat a_p^\dagger \hat a_q
\;+\; \frac{1}{2}\sum_{pqrs} (pq|rs)\,\hat a_p^\dagger \hat a_q^\dagger \hat a_s \hat a_r ,
\label{eq:sqd_secondqH}
\end{equation}
where $h_{pq}$ and $(pq|rs)$ are one- and two-electron integrals in the chosen active orbitals and $E_0$ collects constant contributions. For quantum execution, the active-space Hamiltonian can be mapped to qubit operators via a fermion-to-qubit transformation such as Jordan--Wigner, yielding a qubit Hamiltonian expressed as a linear combination of Pauli strings,
\begin{equation}
\hat H \;=\; \sum_{\ell} c_\ell \hat P_\ell ,
\label{eq:sqd_pauliH}
\end{equation}
with real coefficients $c_\ell$ and Pauli operators $\hat P_\ell$. In the SQD workflow used here, the quantum processor is employed to generate physically relevant determinants by measurement sampling, while Hamiltonian matrix elements are evaluated classically from the underlying molecular integrals.

At each geometry, an approximate correlated state $|\Psi(\boldsymbol{\theta})\rangle$ is prepared on the quantum device and measured in the computational basis to obtain a set of bitstrings,
\begin{equation}
\tilde{\chi} \;=\; \{x_j\}_{x \sim \tilde{p}(x)} ,
\label{eq:sqd_samples}
\end{equation}
where each bitstring $x\in\{0,1\}^{2M}$ encodes the occupation of the $2M$ active spin orbitals and therefore corresponds to a Slater determinant $|x\rangle$ in the active-space determinant basis. In practice, hardware noise can yield samples that violate exact symmetries, most notably the target electron number. To enforce physically valid subspaces, we apply a self-consistent configuration recovery protocol (S-CORE) in which the measured strings are partitioned into $K$ batches, and each batch $b$ is transformed into a recovered determinant subspace $S^{(b)}$ that satisfies the desired electron number (and any additional symmetries imposed). In this work, SQD is used as a classical post-processing procedure applied to computational-basis measurement outcomes. Each measured bitstring is interpreted as a Slater determinant (an electronic configuration) in the active-space occupation-number basis. Given a pool of measured bitstrings for a geometry, SQD constructs one or more recovered determinant subspaces from the unique configurations observed and then performs classical diagonalization of the Hamiltonian projected into each recovered subspace to obtain an approximate ground-state energy and wave function in that subspace.~\cite{robledo2025chemistry,kanno2023quantum}
For each recovered subspace, we form the projected Hamiltonian
\begin{equation}
\hat H^{S^{(b)}} \;=\; \hat P_{S^{(b)}} \hat H \hat P_{S^{(b)}}, \qquad
\hat P_{S^{(b)}} \;=\; \sum_{x \in S^{(b)}} |x\rangle\langle x| ,
\label{eq:sqd_projectedH}
\end{equation}
and obtain its lowest eigenpair by classical diagonalization,
\begin{equation}
\hat H^{S^{(b)}} |\psi^{(b)}\rangle \;=\; E^{(b)} |\psi^{(b)}\rangle .
\label{eq:sqd_subspaceeig}
\end{equation}
The SQD energy at that geometry is taken as the minimum over batches, $E_{\mathrm{SQD}}=\min_b E^{(b)}$. The recovered subspaces are refined self-consistently by updating the estimated orbital occupations from the subspace eigenvectors,
\begin{equation}
n_p \;=\; \frac{1}{K}\sum_{b=1}^{K}\langle \psi^{(b)}|\hat n_p|\psi^{(b)}\rangle ,
\label{eq:sqd_occupdate}
\end{equation}
and iterating the recovery and diagonalization steps for a fixed number of S-CORE iterations (two iterations were used in all production calculations of this work, with K=10 batches per iteration; see Section 2.5 below for full settings).

State preparation on hardware employs a locally unentangled coupled-cluster--Jastrow (LUCJ) circuit, and sampling is performed on the \texttt{ibm\_cleveland} superconducting backend. LUCJ circuit execution is followed by Sample-based Quantum Diagonalization (SQD) post-processing of the measured bitstrings to recover compact determinant subspaces. Recent work has also explored SQD as an approximate solver inside fragment-based multireference frameworks (LASSQD), enabling larger fragment active spaces while retaining near-chemical accuracy.\cite{wang2025sample} To quantify the trade-off between accuracy and quantum cost, we vary the SQD sampling and batching protocol described below. Two error-mitigation strategies are applied during circuit execution. Pauli twirling is enabled on the two-qubit ECR gates to randomize coherent errors into stochastic Pauli noise; twirling on measurement is not applied. Dynamical decoupling sequences are inserted in idle qubit windows. Following measurement, two physical-symmetry constraints are imposed on each sampled bitstring during S-CORE configuration recovery: (i) the correct $\alpha$-spin and $\beta$-spin Hamming weights ($n_\alpha=n_\beta=5$ for the 10-electron active spaces studied here, enforced separately for each spin sector), and (ii) a spin-singlet projection ($S^2=0$). Bitstrings violating either constraint are discarded from the recovered subspace.

\subsection{LUCJ and SQD Simulations}
\label{subsec:lucj_sqd}

\begin{figure}[htbp]
  \centering
  \includegraphics[width=\linewidth]{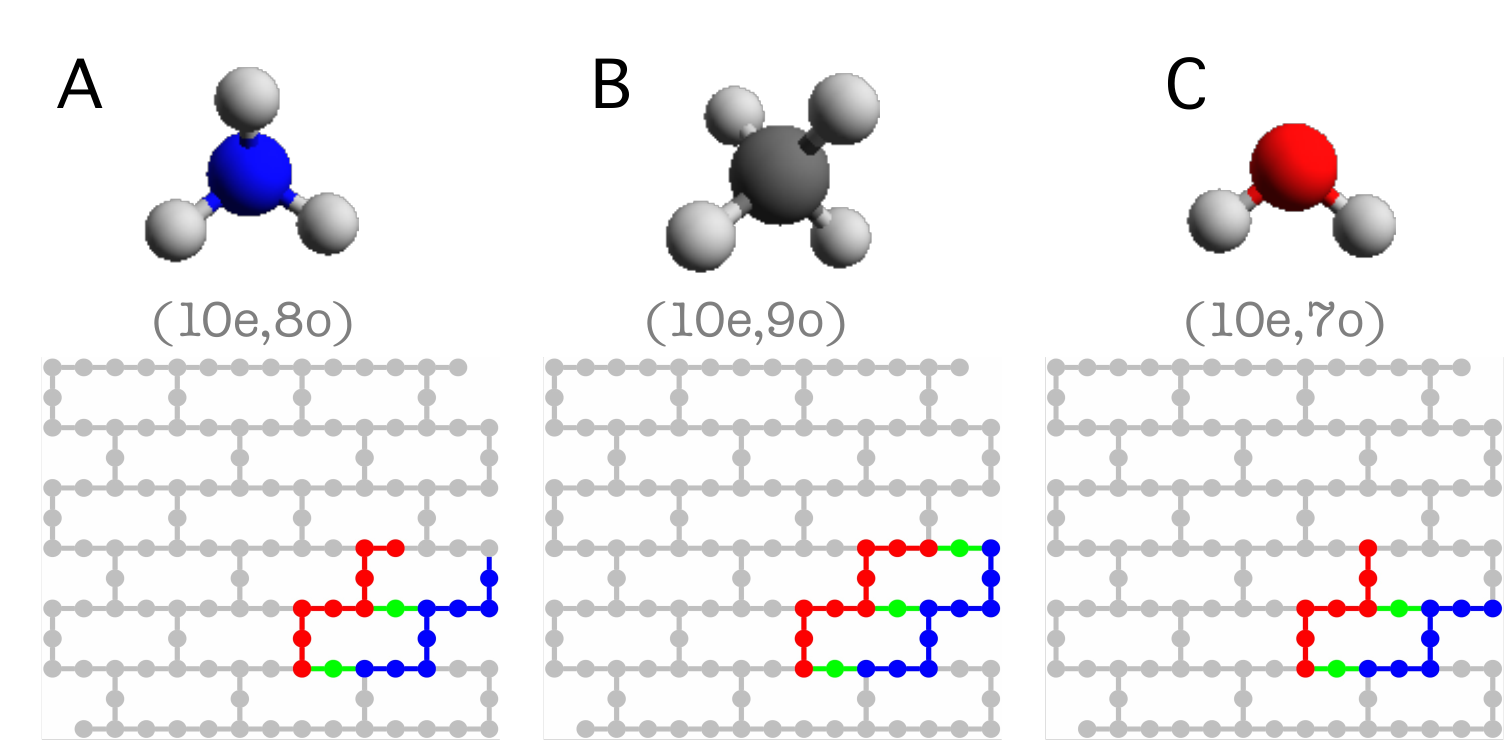}
  \caption{Qubit layouts of the LUCJ circuits executed on \texttt{ibm\_cleveland} for (A) \ce{NH3} (10e,8o; 18 qubits), (B) \ce{CH4} (10e,9o; 21 qubits), and (C) \ce{H2O} (10e,7o; 16 qubits). The device connectivity is shown in gray. Qubits used to encode spin-up and spin-down occupations are indicated in red and blue, respectively, and ancilla qubits are shown in green. The molecular structures above each layout correspond to the simulated solute; carbon, oxygen, nitrogen, and hydrogen atoms are colored gray, red, blue, and white, respectively.}
  \label{fig:qubit_layouts}
\end{figure}

All LUCJ circuits and SQD workflows were implemented using the Qiskit software stack together with \texttt{ffsim} and the Qiskit SQD add-on (\texttt{qiskit-addon-sqd}).\cite{qiskit-addon-sqd} For each molecular geometry, the active-space electronic Hamiltonian was constructed from PySCF integrals in the chosen basis and mapped to a qubit Hamiltonian using the Jordan--Wigner transformation. LUCJ ansatz circuits were generated using \texttt{ffsim}\cite{ffsim} interfaced with Qiskit and executed through Qiskit IBM Runtime using the Qiskit IBM Runtime sampler primitive on the IBM quantum computer \texttt{ibm\_cleveland}, with the qubit layouts shown in Figure~\ref{fig:qubit_layouts}A--C. The LUCJ circuit parameters were obtained from classical restricted closed-shell CCSD calculations and transferred to the corresponding LUCJ operators for quantum sampling.\cite{motta2023bridging}

For the quantum sampling step, we executed the chemistry-inspired LUCJ circuit on the superconducting quantum processor \texttt{ibm\_cleveland} and collected 10\,000 computational-basis measurement shots per molecular geometry. Each shot yields a bitstring that corresponds to an active-space electronic configuration (Slater determinant). To probe finite-sampling effects while holding the underlying hardware dataset fixed, we generated $K=10$ independent SQD reconstructions by drawing $N=200$, 400, or 800 bitstring samples per batch from the 10\,000-shot pool. Each of the K=10 batches at a given N is drawn uniformly at random from the full 10,000-shot pool; the K batches at a given N are statistically independent of one another but are drawn from the same underlying bitstring pool, so they share unique-determinant content with high probability when N is small relative to the number of distinct bitstrings observed. Each batch $b$ defines a recovered determinant set $S^{(b)}$ consisting of the unique bitstrings observed in that batch, with $|S^{(b)}|\le N$, and yields an SQD estimate $E^{(b)}$ and subspace eigenvector $|\psi^{(b)}\rangle$ via classical diagonalization of the projected Hamiltonian. We report the SQD energy for a geometry as the minimum over batches, $E_{\mathrm{SQD}}=\min_b E^{(b)}$, and we use the corresponding $|\psi^{(b)}\rangle$ for subsequent analyses and gradient evaluation.

The SQD energy at a given geometry was obtained by constructing an effective Hamiltonian in the recovered determinant subspace and solving the resulting eigenvalue problem classically. In practice, we performed two iterations of the self-consistent configuration recovery (S-CORE)\cite{robledo2025chemistry} procedure with 10 independent batches per iteration. In each batch, determinant lists were generated from independent quantum samples, the subspace Hamiltonian was assembled from the same active-space integrals, and the lowest eigenpair was computed to obtain a batch energy. This SQD procedure was repeated at each geometry along the trajectory to produce the electronic energies used in the analyses and in the force evaluation described below.

The optimal qubit layout determined by VF2PostLayout was computed once per molecular system (using $n_\mathrm{reps}{=}1$ as a transpilation proxy) and then reused without modification for all 500 MD steps of each production trajectory; this layout-reuse strategy ensures a consistent hardware mapping across the entire trajectory. Transpiled circuit sizes, two-qubit (ECR) gate counts, and circuit depths for each system are reported in Table~S3 and Figure~S1 of the Supporting Information.

\subsection{Force Evaluation}

Forces on the nuclei are obtained from analytical energy gradients evaluated in PySCF for both FCI and SQD. For the classical reference, these analytical gradients are obtained from PySCF's \texttt{mcscf.CASCI} gradient implementation. The same electronic structure definitions and nuclear geometries are used for FCI and SQD, so that any discrepancies in the forces arise solely from differences in the underlying electronic energies. For SQD, the analytical gradients are evaluated in PySCF using reduced density matrices constructed from the selected SQD subspace eigenvector $|\psi^{(b)}\rangle$ at each geometry, so that discrepancies relative to FCI arise from the SQD subspace approximation (and finite-sampling noise) rather than from the force machinery itself. Analytic nuclear gradients within the SQD/ext-SQD framework were previously implemented and demonstrated in our quantum-centric book-ending free-energy workflow, where they were incorporated intermittently into QM/MM propagation.\cite{bazayeva2025quantum}

In the gas-phase tests, we also monitor the root-mean-square (RMS) gradient magnitude as a function of molecular dynamics (MD) step, which allows a direct comparison of force profiles between FCI and SQD. In the quantum AIMD runs, these SQD-derived forces are combined with classical MM forces on the solvent to propagate the nuclei according to the chosen integration scheme.

\subsection{Molecular Dynamics Protocol}

Gas-phase trajectories for \ce{NH3} were propagated in the microcanonical (NVE) ensemble. For validation, we performed a 25~fs trajectory using a 0.5~fs time step (50 MD steps). This trajectory length is intentionally short: it is sufficient to resolve high-frequency intramolecular vibrations of \ce{NH3} (umbrella inversion at ~950 cm-1 has a period of ~35 fs and N-H stretches of ~3300 cm-1 have periods near 10 fs) while keeping the QPU sampling budget tractable for a 50-geometry pointwise SQD-vs-FCI comparison. The trajectory is not intended to sample equilibrium phase-space behavior, only to test pointwise agreement of forces and energy fluctuations. Along this trajectory, FCI and SQD energies and forces were evaluated at every geometry, with FCI serving as the reference for assessing SQD agreement.

For explicit-solvent simulations, QM/MM trajectories of 0.25~ps were generated for \ce{NH3}, \ce{CH4}, and \ce{H2O} in water, again using a time step of 0.5~fs (500 MD steps). For SQD-driven QM/MM dynamics, this corresponds to 500 quantum evaluations of the LUCJ circuit on hardware (one LUCJ execution per MD geometry for each SQD setting) to provide the measurement samples used for SQD energy and gradient evaluation. Initial configurations were equilibrated with a Langevin thermostat (300~K, collision frequency 2.0~ps$^{-1}$): first, 360~ps of NVT MD with gradual heating from 0 to 300~K, followed by 300~ps of NPT MD at 300~K and 1~atm to relax the simulation-box volume. Production NVE runs were initialized from the final NPT restart file, reading both coordinates and velocities directly; FCI and SQD trajectories were thus launched from identical initial conditions to enable direct one-to-one comparisons.

\subsection{Analysis}
\label{sec:analysis}

To quantify SQD accuracy, we compare SQD and FCI results point-by-point along time-aligned trajectories. For energies, we report the mean absolute deviation (MAD) and root-mean-square deviation (RMSD) of $E_{\mathrm{SQD}}(t)-E_{\mathrm{FCI}}(t)$, together with the Pearson correlation coefficient between the two series. Because dynamics is controlled primarily by changes along the sampled potential-energy surface, we also analyze energy-fluctuation agreement using $\Delta E(t)=E(t)-E(0)$ and compute the same statistics for the corresponding $\Delta E$ traces.

For forces, we evaluate stepwise agreement at the gradient level by computing the RMSD between SQD and FCI \emph{analytical} nuclear gradients at each MD frame, and we additionally monitor the RMS gradient magnitude to summarize force-profile consistency along the trajectory.

In the condensed phase, we compute solute--solvent radial distribution functions $g(r)$ from both FCI-driven and SQD-driven QM/MM trajectories using identical binning and normalization, to assess whether the solvent structure responds consistently when nuclear motion is driven by FCI versus SQD energies and gradients.

\section{Results and Discussion}

We first evaluate SQD as a drop-in electronic-structure engine by benchmarking gas-phase \ce{NH3} energies and gradients against the classical reference along representative MD frames. We then test whether these force estimates yield stable NVE trajectories in vacuum. Finally, we move to condensed phase by coupling SQD to a QM/MM engine for \ce{NH3} in water, analyze structural observables, and demonstrate transferability to \ce{CH4} and \ce{H2O} in water.

\subsection{Gas-phase Benchmarks for \texorpdfstring{\ce{NH3}}{NH3}: Energies and Gradients}

We first benchmark SQD for gas-phase \ce{NH3} in a regime where the electronic structure can be evaluated exactly. In this setup, the reference is a full configuration-interaction solution in the complete STO-3G orbital manifold. SQD is tested at three batch sizes, denoted SQD200, SQD400, and SQD800, which correspond to $N=200$, 400, and 800 bitstring samples per batch in the SQD recovery step. The comparison is performed point-by-point over 50 MD frames by aligning trajectories in time and evaluating SQD$-$FCI difference series for both energies and gradients.

\begin{figure}
  \centering
  \includegraphics[width=\linewidth]{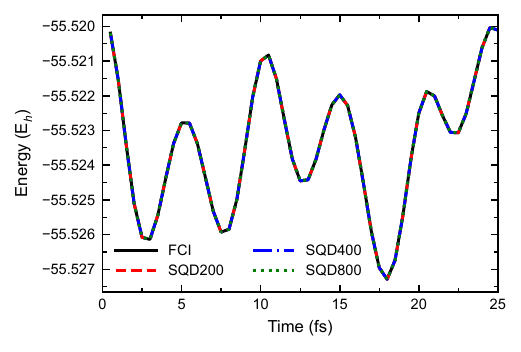}
  \caption{Total electronic energy along the gas-phase \ce{NH3} trajectory in vacuum obtained with FCI/STO-3G reference and SQD at three SQD batch sizes (SQD200, SQD400, SQD800). The SQD curves closely overlay the FCI reference, indicating that SQD reproduces the same potential-energy surface along the sampled path.}
  \label{fig:nh3vac_energy_overlay}
\end{figure}

Figure~\ref{fig:nh3vac_energy_overlay} shows the total electronic energy along this trajectory: the SQD curves track the FCI reference at all three batch sizes, confirming that the sampled subspace reproduces the same potential-energy surface.

\paragraph{Force errors and energy-fluctuation agreement.}
For the gas-phase \ce{NH3} benchmark, we assess SQD as a drop-in electronic-structure engine for dynamics by quantifying (i) \emph{force/gradient errors}, which directly control trajectory propagation, and (ii) \emph{energy-fluctuation agreement}, which tests whether SQD reproduces the same step-to-step changes along the sampled potential-energy surface. Energy agreement is therefore evaluated in the form most relevant to dynamics, namely fluctuations relative to the initial frame, $\Delta E(t)=E(t)-E(t_0)$, rather than absolute energy offsets.

The most striking result is the near-zero and extremely small error in the RMS gradient magnitude. Across the 50-step trajectory, the RMS-gradient error $g_{\mathrm{SQD}}(t)-g_{\mathrm{FCI}}(t)$ is centered essentially at zero (mean $\approx 6\times 10^{-8}$ a.u.) with mean absolute error (MAE) of $2.52\times 10^{-7}$ a.u. (SQD200), $2.53\times 10^{-7}$ a.u. (SQD400), and $2.53\times 10^{-7}$ a.u. (SQD800). The corresponding RMSE values are $3.41\times 10^{-7}$ a.u. (SQD200) and $3.42\times 10^{-7}$ a.u. (SQD400/SQD800), with worst-case deviations below $9.33\times 10^{-7}$ a.u. over the full trajectory. These statistics indicate that SQD reproduces not only the overall force scale but also the frame-to-frame variations needed for stable propagation, with no evidence of a systematic bias in the gradient magnitude. Importantly, increasing the SQD batch size from 200 to 400 and 800 produces no meaningful change in these error metrics within numerical precision, showing that SQD200 is already sufficient to recover FCI-level force behavior for this  molecule in the present regime.

Energy-fluctuation agreement is similarly strong. The fluctuation-level error $\Delta E_{\mathrm{SQD}}(t)-\Delta E_{\mathrm{FCI}}(t)$ remains extremely small throughout the trajectory, with MAE of $1.38\times 10^{-7}$~$E_h$ (SQD200) and $1.35\times 10^{-7}$~$E_h$ (SQD400/SQD800), RMSE of $(1.72\text{--}1.73)\times 10^{-7}$~$E_h$, and a maximum absolute deviation of $4.2\times 10^{-7}$~$E_h$. Consistent with this, the net change in the fluctuation signal over the trajectory window is reproduced essentially exactly: $\Delta E_{\mathrm{FCI}}(t_{\mathrm{final}})=8.7\times 10^{-5}$~$E_h$ and $\Delta E_{\mathrm{SQD}}(t_{\mathrm{final}})=8.7\times 10^{-5}$~$E_h$ for all three batch sizes, with a residual drift mismatch of only $1\times 10^{-7}$~$E_h$.

We note that the gradient error statistics reported here approach the numerical-precision floor of the double-precision FCI gradient evaluation in PySCF. Differences at this magnitude should therefore be interpreted as ``indistinguishable from the classical reference'' rather than as physically meaningful residuals; the meaningful conclusion is that SQD200 already saturates the achievable agreement with FCI for vacuum NH3 at the present basis-set/active-space resolution.

Together, the force-error statistics and the $\Delta E$ diagnostics show that SQD matches FCI for both gradients and MD-relevant energy changes along the gas-phase \ce{NH3} trajectory under these conditions. The key practical point is that the smallest batch size tested, SQD200, already attains the converged behavior observed at larger batch sizes. In other words, SQD is able to recover FCI quality along a dynamical path while retaining only a compact, data-driven subset of configurations, underscoring the effectiveness of SQD as a configuration-recovery and dynamics-capable quantum workflow.

\subsection{Stability of SQD-driven MD Trajectories in Vacuum}

Using the same gas-phase \ce{NH3} system, we next examine whether SQD can function as a stable force engine for microcanonical (NVE) propagation in vacuum. In this context, the relevant concern is not only pointwise agreement at isolated geometries, but whether finite-sampling and hardware noise lead to systematic bias or step-to-step irregularities in forces that would manifest as unstable dynamics. We therefore monitor the RMS nuclear gradient magnitude along the trajectory and interpret stability in conjunction with the fluctuation-level energy agreement reported above for $\Delta E(t)$.

\begin{figure}
  \centering
  \includegraphics[width=\linewidth]{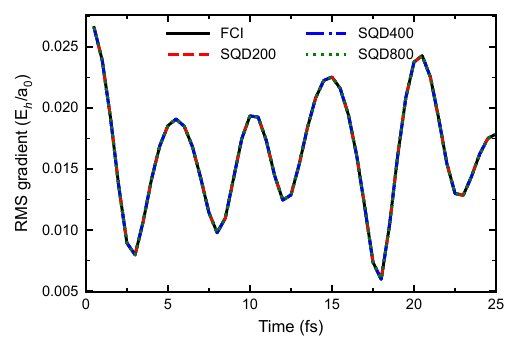}
  \caption{Root-mean-square (RMS) nuclear gradient magnitude along the gas-phase \ce{NH3} trajectory in vacuum obtained with FCI/STO-3G reference and SQD at three SQD batch sizes (SQD200, SQD400, SQD800). The SQD curves overlay the FCI reference across the full trajectory, indicating that SQD reproduces the force scale and its temporal modulation along the NVE path.}
  \label{fig:nh3vac_rmsgrad_overlay}
\end{figure}

As shown in Figure~\ref{fig:nh3vac_rmsgrad_overlay}, the RMS gradient magnitude obtained from SQD closely overlays the FCI reference throughout the trajectory for all three batch sizes. This qualitative agreement is reinforced by the quantitative error statistics in the preceding subsection, where the RMS-gradient difference $g_{\mathrm{SQD}}(t)-g_{\mathrm{FCI}}(t)$ is centered near zero with sub-$10^{-6}$ a.u. worst-case deviations and indistinguishable MAE and RMSE values for SQD200, SQD400, and SQD800. Together with the essentially exact reproduction of the fluctuation signal $\Delta E(t)$, these results show no evidence of systematic force bias or accumulated mismatch over the 50-step window. Practically, the absence of any measurable improvement when increasing the batch size beyond SQD200 indicates that the cheapest setting already delivers stable, FCI-consistent force behavior for this vacuum \ce{NH3} benchmark.

Beyond pointwise gradient agreement, the defining diagnostic for NVE stability is conservation of the total energy $E_\mathrm{total} = T_\mathrm{kinetic} + V_\mathrm{potential}$ along the trajectory. We monitored $E_\mathrm{total}$ along the FCI- and SQD-driven 25~fs vacuum \ce{NH3} NVE trajectories (Figure~S8, Supporting Information), all four of which use the same 0.5~fs timestep and start from identical coordinates and velocities. All four exhibit the same oscillation, with a peak-to-peak amplitude of 0.1208~mE$_\mathrm{h}$, and the same net endpoint change over the window, $-0.0099$~mE$_\mathrm{h}$ for FCI and $-0.0097$~mE$_\mathrm{h}$ for each SQD batch size. This common oscillation is consistent with finite-timestep integration of a light hydride without bond constraints and is a property of the integrator rather than of the electronic-structure method. Against that shared background, the quantity that measures the electronic-structure difference is the separation between the SQD and FCI curves, which does not exceed $4.8\times10^{-4}$~mE$_\mathrm{h}$ at any point for any batch size. Quantum-hardware noise therefore introduces no anomalous dissipation or energy injection relative to the classical reference propagated from the same initial conditions. Because the window is only 25~fs, this diagnostic establishes short-time energy conservation; it does not establish long-time NVE stability or converged thermodynamic sampling.

\subsection{\texorpdfstring{\ce{NH3}}{NH3} in Water: QM/MM Energies, Analytical Gradients, and Structural Observables}

The condensed-phase setting provides a more stringent test of SQD because the solvent continuously perturbs the QM region, and short-time dynamics depend on instantaneous forces and on physically meaningful energy fluctuations rather than absolute energy offsets. We therefore benchmark aqueous \ce{NH3} in a QM/MM environment by comparing SQD and FCI point-by-point over an aligned 500-frame trajectory segment ($\Delta t = 0.5$~fs, 250~fs total), where FCI provides an exact reference (FCI). As defined in Section~\ref{sec:analysis}, we focus on two MD-relevant diagnostics: (i) RMS-gradient-magnitude errors and (ii) fluctuation-level energy agreement using $\Delta E(t)=E(t)-E(0)$.

\textbf{Force and gradient fidelity in water.}
Across the aqueous trajectory, SQD reproduces FCI forces with errors that remain centered essentially at zero and extremely small over time. For all three batch sizes (SQD200, SQD400, and SQD800), the RMS-gradient magnitude error distribution exhibits a near-zero mean of $+4.0\times 10^{-6}\ E_h/a_0$, with MAE $2.3\times 10^{-5}\ E_h/a_0$ (0.0268~kcal~mol$^{-1}$~\AA$^{-1}$), RMSE $5.0\times 10^{-5}\ E_h/a_0$ (0.0596~kcal~mol$^{-1}$~\AA$^{-1}$), and a maximum absolute deviation of $3.59\times 10^{-4}\ E_h/a_0$ (0.426~kcal~mol$^{-1}$~\AA$^{-1}$). The frame-by-frame distribution is strongly concentrated near zero. For SQD200, the median absolute gradient error is 0.0034~kcal~mol$^{-1}$~\AA$^{-1}$, and 95\% and 99\% of frames fall below 0.15 and 0.26~kcal~mol$^{-1}$~\AA$^{-1}$, respectively. Moreover, 98.2\% of frames satisfy $|g_{\mathrm{SQD}}-g_{\mathrm{FCI}}| < 0.2$~kcal~mol$^{-1}$~\AA$^{-1}$ and all frames satisfy $|g_{\mathrm{SQD}}-g_{\mathrm{FCI}}| < 0.5$~kcal~mol$^{-1}$~\AA$^{-1}$.

\begin{figure}[htbp]
  \centering
  \includegraphics[width=\columnwidth]{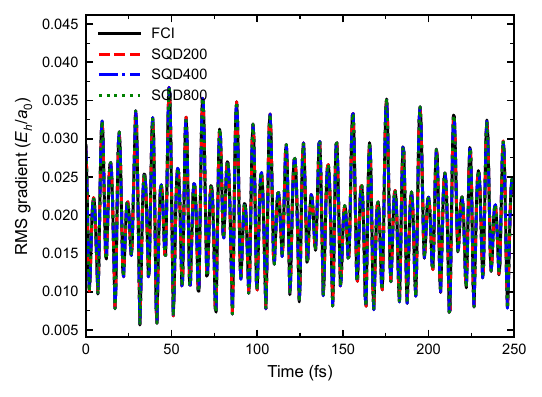}
  \caption{Root-mean-square (RMS) gradient magnitude ($E_h/a_0$) along the aqueous-phase molecular dynamics trajectory of NH$_3$ in explicit water, comparing classical FCI with SQD using batch sizes $N = 200$, 400, and 800 samples per batch. Gradients are plotted versus simulation time (fs) with $\Delta t = 0.5$ fs. SQD reproduces the FCI RMS-gradient profile closely for all batch sizes, indicating consistent force accuracy in solvent.}
  \label{fig:nh3_aq_rmsgrad_overlay}
\end{figure}

A useful normalization is to compare these errors to the characteristic force scale in water. Along this segment, the mean FCI RMS-gradient magnitude is 0.02018~$E_h/a_0$ (23.9~kcal~mol$^{-1}$~\AA$^{-1}$). The SQD200 gradient MAE of 0.0268~kcal~mol$^{-1}$~\AA$^{-1}$ is only 0.11\% of that mean magnitude, and the RMSE of 0.0596~kcal~mol$^{-1}$~\AA$^{-1}$ is only 0.25\%. Relative to intrinsic solvent-induced variability, the standard deviation of the FCI RMS gradients is 0.00677~$E_h/a_0$ (8.03~kcal~mol$^{-1}$~\AA$^{-1}$), while the SQD gradient RMSE corresponds to only 0.74\% of that variability. These results indicate that the SQD force perturbation is far smaller than the physical fluctuations imposed by the aqueous environment. Correlation diagnostics corroborate this conclusion, with Pearson $r = 0.9999$ between SQD and FCI RMS-gradient magnitudes across the trajectory. The resulting close overlap in the RMS-gradient profiles is shown in Figure~\ref{fig:nh3_aq_rmsgrad_overlay}.

\textbf{Energy-fluctuation agreement.}
For stability, the most revealing energy diagnostic is whether SQD reproduces \emph{step-to-step energy changes} along the sampled potential-energy surface. The $\Delta E$ error series is centered near zero for all batch sizes, with mean $\Delta E$ error of $-1.0\times 10^{-6}\ E_h$, MAE $7.8\times 10^{-5}\ E_h$ (0.0491~kcal~mol$^{-1}$), RMSE $3.46\times 10^{-4}\ E_h$ (0.217~kcal~mol$^{-1}$), and a maximum absolute deviation of $2.717\times 10^{-3}\ E_h$ (1.705~kcal~mol$^{-1}$). For SQD200, the median absolute $\Delta E$ error is 0.0015~kcal~mol$^{-1}$, while 95\% and 99\% of steps fall below 0.279 and 1.40~kcal~mol$^{-1}$, respectively. In addition, 92.6\% of steps satisfy $|\Delta E_{\mathrm{SQD}}-\Delta E_{\mathrm{FCI}}| < 0.1$~kcal~mol$^{-1}$, 97.0\% satisfy $<0.5$~kcal~mol$^{-1}$, and 98.2\% satisfy $<1.0$~kcal~mol$^{-1}$.

\begin{figure}[htbp]
  \centering
  \includegraphics[width=\columnwidth]{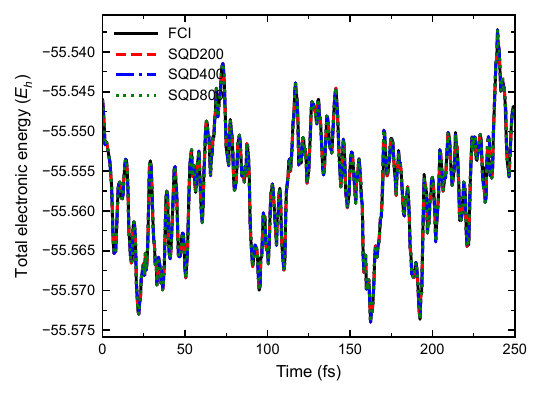}
  \caption{Total electronic energy ($E_h$) along the aqueous-phase molecular dynamics trajectory of NH$_3$ in explicit water, comparing classical FCI with SQD using batch sizes $N = 200$, 400, and 800 samples per batch. Energies are shown as a function of simulation time (fs) with $\Delta t = 0.5$ fs. The SQD and FCI energy traces are in excellent agreement throughout the trajectory.}
  \label{fig:nh3_aq_energy_overlay}
\end{figure}

To contextualize these errors against the intrinsic condensed-phase energy variability, the FCI step-to-step energy changes have a standard deviation of 0.00206~$E_h$ (1.29~kcal~mol$^{-1}$), whereas the SQD $\Delta E$ RMSE is 0.000347~$E_h$ (0.217~kcal~mol$^{-1}$), which is 16.8\% of the natural FCI fluctuation scale. Correlation diagnostics are consistent with strong fluctuation agreement: the total energies have Pearson $r = 0.9990$ between SQD and FCI, and the step-to-step energy changes have Pearson $r = 0.9859$ between $\Delta E_{\mathrm{SQD}}$ and $\Delta E_{\mathrm{FCI}}$. The resulting close overlap in the total electronic-energy traces is shown in Figure~\ref{fig:nh3_aq_energy_overlay}.

While absolute energy conservation is not expected in QM/MM and absolute offsets can be non-informative, the net drift mismatch over the segment provides a coarse check for systematic divergence. Over 250~fs, FCI changes by $-0.001584\ E_h$ ($-0.994$~kcal~mol$^{-1}$), whereas SQD changes by approximately $-0.00211\ E_h$ ($-1.32$~kcal~mol$^{-1}$). The resulting drift mismatch is only $(-5.24$ to $-5.28)\times 10^{-4}\ E_h$, corresponding to $-0.329$ to $-0.331$~kcal~mol$^{-1}$. This small mismatch, together with the near-zero mean $\Delta E$ error, argues against spurious heating artifacts introduced by SQD in the aqueous environment over this time window.

\textbf{SQD delivers reliable energies and gradients with a small recovered subspace.}
A central practical outcome is that SQD200, SQD400, and SQD800 are statistically indistinguishable on the MD-relevant metrics reported here. For gradients, all three batch sizes yield MAE 0.0268~kcal~mol$^{-1}$~\AA$^{-1}$ and RMSE 0.0596~kcal~mol$^{-1}$~\AA$^{-1}$ with identical tail behavior (maximum 0.426~kcal~mol$^{-1}$~\AA$^{-1}$). For energy fluctuations, all three batch sizes yield $\Delta E$ MAE 0.0491~kcal~mol$^{-1}$ and $\Delta E$ RMSE 0.217~kcal~mol$^{-1}$, with the same worst-case deviation of 1.705~kcal~mol$^{-1}$ and unchanged correlations with FCI. This saturation indicates that 200 samples (determinants) per SQD batch already capture the dominant configurations required to reproduce FCI quality forces and energy fluctuations for aqueous \ce{NH3} at this level of theory. The 400 and 800 batch sizes are therefore best interpreted as benchmarking controls that confirm the robustness of SQD200 rather than necessary operating points for stable and accurate QM/MM propagation.

\begin{figure}[htbp]
  \centering
    \includegraphics[width=\linewidth]{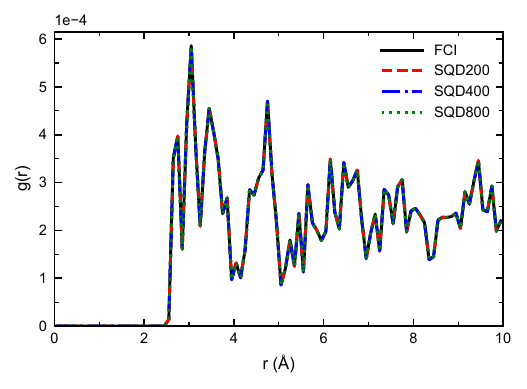}
  \caption{Nitrogen-centered solute--solvent radial distribution function, $g(r)$, between \ce{NH3} atom N1 and water oxygen from the QM/MM trajectory in water ($\Delta t=0.5$ fs), comparing FCI with SQD at 200, 400, and 800 samples per batch. SQD reproduces the FCI solvation-shell structure, including the first-shell maximum at $r=3.05~\mathrm{\AA}$, the first-shell minimum at $r=3.95~\mathrm{\AA}$, and the second-shell maximum at $r=4.75~\mathrm{\AA}$.}
  \label{fig:rdf_nh3_aq}
\end{figure}

\paragraph{QM/MM structural observables: SQD preserves the local solvation structure around \texorpdfstring{\ce{NH3}}{NH3}.}
A stringent condensed-phase check is whether SQD-driven QM/MM dynamics reproduces the same solute--solvent organization as the FCI reference. We therefore compare the nitrogen-centered solute--solvent radial distribution function, $g(r)$, between \ce{NH3} atom N1 and water oxygen (Figure~\ref{fig:rdf_nh3_aq}). The SQD and FCI curves overlay essentially perfectly across the full distance range, reproducing the same first-shell and second-shell features. In particular, the first-shell maximum occurs at $r=3.05~\mathrm{\AA}$ for FCI and is reproduced at the same position by SQD, followed by a first-shell minimum at $r=3.95~\mathrm{\AA}$ and a second-shell maximum at $r=4.75~\mathrm{\AA}$. Quantitatively, the binwise differences relative to FCI remain at the $10^{-6}$ level in $g(r)$ (maximum absolute deviation $\approx 1.0\times 10^{-6}$; MAE $\approx 1.4\times 10^{-7}$; Pearson $r \approx 0.9999$), confirming that SQD preserves the short-range hydration structure around \ce{NH3} at a level consistent with the energy and force agreement.

These aqueous QM/MM benchmarks reinforce the same conclusion reached in the gas-phase and NVE tests: SQD reproduces FCI-quality forces and the physically relevant energy fluctuations in explicit solvent while preserving local solvation structure. In practice, the smallest batch size ($N=200$) already saturates on the MD-relevant metrics reported here for aqueous NH3, and higher settings mainly serve as robustness checks. For systems where larger active spaces demand more sampling, SQD responds efficiently: CH4(aq) shows a 600$\times$ reduction in gradient error from SQD200 to SQD400, reaching near-FCI accuracy, demonstrating that the method systematically improves with modest increases in batch size (Section 3.4).

\subsection{Extension to \texorpdfstring{\ce{CH4}}{CH4} and \texorpdfstring{\ce{H2O}}{H2O} in Water}

The \ce{NH3}(aq) results above establish that SQD can be coupled to a standard QM/MM engine and can reproduce the FCI reference (FCI) in both force measures and step-to-step energy fluctuations, while preserving the local solvation structure. We next probe transferability across two chemically distinct aqueous test cases. \ce{CH4} provides a hydrophobic benchmark dominated by nonspecific solute--solvent interactions, whereas \ce{H2O} challenges the method in a strongly hydrogen-bonded environment where small force biases can rapidly perturb local structure. To keep the main manuscript focused on the detailed \ce{NH3}(aq) case, we provide the full energy and gradient overlays and all RDFs for these extensions in the Supporting Information, and summarize here the key quantitative MD metrics extracted from 0.25~ps trajectories (500 frames at 0.5~fs spacing).

\paragraph{\ce{CH4}(aq): force fidelity and energy-fluctuation agreement.}
For \ce{CH4}(aq), SQD tracks the FCI reference extremely closely in both the RMS-gradient magnitude and the energy-fluctuation series. At the lowest batch size (SQD200), the RMS-gradient error $g_{\mathrm{SQD}}(t)-g_{\mathrm{FCI}}(t)$ has MAE $=0.063$~kcal~mol$^{-1}$~\AA$^{-1}$ and RMSE $=0.086$~kcal~mol$^{-1}$~\AA$^{-1}$, with a maximum absolute deviation of $0.326$~kcal~mol$^{-1}$~\AA$^{-1}$ across the 500 frames. These force errors are small on the scale relevant for short-time trajectory propagation and are consistent with SQD acting as a stable electronic-structure backend for hydrophobic hydration dynamics. The corresponding energy-fluctuation error series, defined as $\Delta E_{\mathrm{SQD}}(t)-\Delta E_{\mathrm{FCI}}(t)$ with $\Delta E(t)=E(t)-E(0)$, yields MAE $=0.041$~kcal~mol$^{-1}$ and RMSE $=0.054$~kcal~mol$^{-1}$, with a maximum absolute deviation of $0.192$~kcal~mol$^{-1}$. The endpoint mismatch in the net energy change over 0.25~ps is small ($-0.042$~kcal~mol$^{-1}$ for SQD200), and it is reduced to essentially zero at larger batch sizes (0.00044 and 0.00005~kcal~mol$^{-1}$ for SQD400 and SQD800, respectively). In other words, SQD200 is already within 1 kcal mol$^{-1}$ of the FCI reference for the fluctuation metric, while SQD400 and SQD800 drive the residual finite-sampling noise down by roughly two orders of magnitude (Figures~S2--S3). The gradient-error improvement is particularly pronounced: gradient MAE drops from 0.063 (SQD200) to $5.9\times10^{-4}$ (SQD400) kcal~mol$^{-1}$~\AA$^{-1}$, a reduction of approximately 600$\times$. This system-dependent behavior contrasts with the saturated SQD200 response observed for vacuum and aqueous \ce{NH3}, and indicates that the effective batch size required for SQD convergence depends on the chemistry of the active space and not only on its formal size; we discuss the implications in the Conclusions.

\paragraph{QM/MM structural observables: SQD reproduces hydrophobic hydration around \texorpdfstring{\ce{CH4}}{CH4}.}
After the polar \ce{NH3} test case, \ce{CH4} provides a complementary structural benchmark because its hydration structure is subtle and arises primarily from solvent packing rather than strong directional solute--solvent interactions. We therefore analyze the carbon-centered RDF, $g(r)$, between the \ce{CH4} carbon atom C1 and water oxygen. The SQD and FCI curves coincide across the full distance range, indicating that SQD does not introduce spurious structuring or depletion of water around the hydrophobe. The first-shell maximum occurs at $r=3.55~\mathrm{\AA}$ for FCI and is reproduced at the same position by SQD, followed by a first minimum at $r=5.05~\mathrm{\AA}$ and a second-shell maximum at $r=6.25~\mathrm{\AA}$. Quantitatively, the binwise deviations relative to FCI remain at the $10^{-6}$ level in $g(r)$ (for SQD200, maximum absolute deviation $\approx 3.0\times 10^{-6}$; MAE $\approx 5.9\times 10^{-7}$; Pearson $r \approx 0.9999$), while SQD400 and SQD800 are identical to FCI within the reported precision. This agreement confirms that SQD preserves the local solvent packing around \ce{CH4} in the QM/MM setting, consistent with the corresponding force and energy diagnostics.

\paragraph{\ce{H2O}(aq): force fidelity and energy-fluctuation agreement.}
For the \ce{H2O} solute in OPC water, SQD again reproduces the FCI reference with small force errors and tight energy-fluctuation agreement. The RMS-gradient error has MAE $=0.430$~kcal~mol$^{-1}$~\AA$^{-1}$ and RMSE $=1.146$~kcal~mol$^{-1}$~\AA$^{-1}$, with 82.0\% of frames satisfying $|g_{\mathrm{SQD}}-g_{\mathrm{FCI}}|<0.5$~kcal~mol$^{-1}$~\AA$^{-1}$ and 89.2\% satisfying $<1.0$~kcal~mol$^{-1}$~\AA$^{-1}$. The energy-fluctuation error series yields MAE $=0.139$~kcal~mol$^{-1}$ and RMSE $=0.316$~kcal~mol$^{-1}$, with 90.8\% of frames satisfying $|\Delta E_{\mathrm{SQD}}-\Delta E_{\mathrm{FCI}}|<0.5$~kcal~mol$^{-1}$ and 96.6\% satisfying $<1.0$~kcal~mol$^{-1}$. The endpoint mismatch in the net $\Delta E$ change over 0.25~ps is $-0.355$~kcal~mol$^{-1}$. Within the resolution of these metrics, SQD200, SQD400, and SQD800 are statistically indistinguishable for this trajectory, indicating that the accuracy has effectively saturated at the lowest batch size explored here. In fact, the metrics for SQD200, SQD400, and SQD800 are numerically identical to the precision reported in Table~S6, not merely statistically indistinguishable. We attribute this exact equality to the structure of the underlying 10,000-shot bitstring pool: for the small (10e,7o) active space of \ce{H2O}, the dominant low-energy configurations comprise a tightly limited set of unique determinants, so that any subsample of $N=200$, 400, or 800 bitstrings drawn from the same pool selects essentially the same set of unique configurations after symmetry filtering. The recovered subspace $\mathcal{S}_b$ is therefore nearly identical at all three batch sizes, and the projected-Hamiltonian eigenvalues coincide.

\paragraph{QM/MM structural observables: SQD maintains the short-range oxygen--oxygen structure for \texorpdfstring{\ce{H2O}}{H2O} in water.}
For \ce{H2O} as a polar solute, local structure is dominated by the hydrogen-bond network, so agreement with the FCI reference requires not only similar radial packing but also preservation of the characteristic first- and second-shell oxygen correlations. We therefore examine the O1--O$_\mathrm{w}$ radial distribution function, $g(r)$, between the solute oxygen atom O1 and water oxygen. The SQD and FCI curves coincide across the full distance range, capturing the same pronounced first-shell peak and subsequent shell structure. The first maximum appears at $r=3.15~\mathrm{\AA}$ for FCI and is reproduced by SQD, followed by a first minimum at $r=4.35~\mathrm{\AA}$ and a second-shell maximum at $r=5.45~\mathrm{\AA}$. Quantitatively, the RDF differences relative to FCI remain at the $10^{-6}$ level (for SQD200, maximum absolute deviation $\approx 3.0\times 10^{-6}$; MAE $\approx 5.9\times 10^{-7}$; Pearson $r \approx 0.9999$), while SQD400 and SQD800 are indistinguishable from FCI within the precision of the output. This close agreement indicates that SQD preserves the local oxygen coordination environment and the near-neighbor structure of liquid water in the QM/MM setting.

Overall, the \ce{CH4} and \ce{H2O} extensions support the main \ce{NH3}(aq) conclusion: within these minimal-basis active spaces and short-time QM/MM windows, SQD provides a transferable electronic-structure backend that preserves both MD-relevant forces and condensed-phase structure. The CH4 case further shows that while SQD200 is already chemically accurate for fluctuation metrics, increasing $N$ can systematically suppress the remaining finite-sampling noise when desired for benchmarking.

\subsection{SQD as a Practical Quantum Backend for Condensed-Phase AIMD}

Across the gas-phase and QM/MM benchmarks, SQD shows why it matters for quantum computing in chemistry: a real, noisy quantum processor can act as an electronic-structure backend inside a standard AIMD workflow while still recovering FCI-level behavior for small solvated molecules. In the minimal STO-3G settings studied here, SQD reproduces FCI energies and force-driven nuclear dynamics to within 1 kcal mol$^{-1}$ of the FCI reference at the level of mean absolute errors (sub-kcal~mol$^{-1}$ MAE for energy fluctuations and sub-kcal~mol$^{-1}$~\AA$^{-1}$ MAE for gradients across all three solutes), and it yields condensed-phase trajectories whose local structure and short-time solvation signatures are statistically indistinguishable from their classical references, even at modest quantum shot counts. That combination is the key point: SQD is not merely producing reasonable energies in isolation, but delivering stable, usable forces in a dynamical context, which is the real bottleneck for turning quantum algorithms into predictive chemical simulation tools. This establishes a concrete baseline that quantum hardware can already be wired into realistic condensed-phase simulation pipelines, and it reframes the next challenges as engineering targets rather than abstract promises: reducing instantaneous noise without brute-force shots, improving force estimators beyond fragile finite differences, and scaling to larger active spaces, stronger multireference character, and more chemically realistic basis sets across hardware platforms with different error profiles.

\section{Conclusions}

In this work, we demonstrate QM/MM molecular dynamics driven by a quantum electronic-structure engine through a quantum classical workflow that combines chemistry-inspired LUCJ circuit execution with sample-based quantum diagonalization (SQD). Using quantum-generated bitstring samples to recover compact determinant subspaces, SQD delivers FCI-quality energies and analytical gradients in the complete STO-3G manifolds, enabling a point-by-point assessment of MD-relevant errors and a direct drop-in replacement of the classical QM engine in a standard QM/MM MD driver. Across gas-phase \ce{NH3} benchmarks, SQD reproduces FCI energy fluctuations and force profiles with excellent fidelity, and SQD-driven NVE trajectories remain stable with indistinguishable behavior at the level resolved by the simulations. In explicit OPC water, the same LUCJ+SQD approach preserves this agreement for QM/MM energies and forces and yields solute--solvent structural observables, including radial distribution functions, that closely track the FCI reference.
The convergence behavior of the SQD-driven dynamics with respect to batch size is system dependent. For NH3 in vacuum, NH3(aq), and H2O(aq), the smallest batch size tested (SQD200) already saturates the MD-relevant accuracy metrics; in the H2O(aq) case the recovered determinant subspace itself is essentially identical at SQD200, SQD400, and SQD800, reflecting the small unique-determinant pool generated by the (10e,7o) active space (see Section 3.4 and SI). For CH4(aq), increasing the batch size from SQD200 to SQD400 yields a $\sim$600$\times$ reduction in gradient error, reaching near-FCI accuracy -- a demonstration that SQD convergence can be systematically and efficiently improved by modest increases in sampling. In practice, we therefore recommend that SQD batch-size convergence be confirmed on a per-system basis rather than assumed from a single benchmark. Extensions to \ce{CH4}(aq) and \ce{H2O}(aq) further support the generality of the approach, establishing LUCJ+SQD as an efficient and practical route for performing quantum-enabled QM/MM molecular dynamics in solution.

Within the broader vision of quantum-centric supercomputing for chemistry,\cite{robledo2025chemistry,alexeev2024quantum,alexeev2025perspective} the present results contribute a concrete milestone: an SQD-based CI solver seamlessly integrated into a standard QM/MM AIMD driver delivers FCI-quality energies and forces along a 250~fs trajectory using superconducting quantum hardware. This demonstrates that quantum processors are already capable of driving chemically meaningful molecular dynamics in explicit solvent--a significant advance that establishes quantum hardware as a practical and reliable engine for predictive chemical simulation. The natural extensions of this roadmap are actively being pursued: active-space embedding via DMET-SQD\cite{shajan2025toward} enables quantum accuracy for extended molecules, and the fragment-based EWF framework\cite{shajan2025molecular}--demonstrated at the single-point level for protein-scale systems--provides a direct path toward QM/MM molecular dynamics with analytical gradient implementations for chemically realistic, large-scale environments. Combined with quantum-hardware-driven free energy perturbation\cite{li2026fep} these directions collectively position the present AIMD workflow as a foundation for quantum-driven simulation across the full range of challenges in computational chemistry and pharmacology. The present demonstration uses the STO-3G minimal basis, short timescale simulations and compact active spaces ((10e,8o) for \ce{NH3}, (10e,9o) for \ce{CH4}, and (10e,7o) for \ce{H2O}), chosen to be tractable on contemporary 16--21 qubit hardware; lifting these constraints via the embedding strategies outlined above is the immediate next step. Beyond increasing the sample budget, subspace augmentation through extended SQD (ext-SQD), in which the dominant determinants of the recovered subspace are supplemented by selected single excitations, offers a route to tighter accuracy at fixed shot count and is a natural next step for quantum-centric AIMD.\cite{barison2025quantum,bazayeva2025quantum}

\section{Code Availability}
Code used in this work is publicly available through open-source repositories and documentation. Qiskit, \texttt{ffsim}, and Qiskit IBM Runtime used for the LUCJ simulations are available at \url{https://github.com/qiskit-community/ffsim}, \url{https://github.com/Qiskit/qiskit}, and \url{https://github.com/Qiskit/qiskit-ibm-runtime}, respectively. The configuration-recovery implementation is distributed as the Qiskit SQD add-on (\url{https://github.com/Qiskit/qiskit-addon-sqd}). Classical electronic-structure calculations were performed with PySCF (\url{https://github.com/pyscf/pyscf}). A tutorial demonstrating the full SQD workflow is available at \url{https://qiskit.github.io/qiskit-addon-sqd/tutorials/01_chemistry_hamiltonian.html}. The results reported here were obtained with Qiskit 2.0.0, qiskit-ibm-runtime 0.37.0, qiskit-addon-sqd 0.9.0, \texttt{ffsim} 0.0.49, PySCF 2.8.0, ray 2.42.1, and AmberTools 23.

\begin{acknowledgement}

We gratefully acknowledge financial support from the National Science Foundation through the CSSI Frameworks program (Grant OAC-2435622) and from the National Institutes of Health (Grant GM130641). We also acknowledge computational resources provided by the Institute for Cyber-Enabled Research High Performance Computing Center (iCER HPCC) at Michigan State University and the high-performance computing facilities at the Cleveland Clinic Foundation. We thank Abdullah Ash Saki for providing code to identify favorable qubit layouts for LUCJ circuits on IBM quantum processors. We further thank Mario Motta and Thaddeus Pellegrini of IBM Quantum for the SQD code, and for useful discussions and technical help. We also thank Dr. Andreas G\"otz (University of California San Diego) for fruitful discussions and suggestions.

\end{acknowledgement}

\begin{suppinfo}

The Supporting Information includes: Section~1 (Extended Computational Details) with subsections on the OPC water model and box setup (Tables~S1--S2), the AMBER MD protocol, LUCJ circuit construction and qubit layout including transpiled circuit resources (Table~S3 and Figure~S1), and the measurement, error-mitigation, and SQD post-processing settings; Section~2 with full energy and gradient overlays for \ce{CH4}(aq) (Figures~S2--S3) and the C1--Ow radial distribution function (Figure~S4); Section~3 with energy and gradient overlays for \ce{H2O}(aq) (Figures~S5--S6) and the O1--Ow radial distribution function (Figure~S7); Section~4 (Tables~S4--S6) collecting all quantitative MD accuracy metrics (gradient and $\Delta E$ mean absolute errors, RMSE, maximum absolute deviations, and endpoint drift) for the \ce{NH3}(aq), \ce{CH4}(aq), and \ce{H2O}(aq) trajectories; and Section~5 (Figure~S8) reporting the $E_\mathrm{total}$ NVE energy-conservation diagnostic for the vacuum \ce{NH3} trajectory; and Section~6 listing the abbreviations used in this work.

\end{suppinfo}

\bibliography{sqd_qmmm}

\end{document}